\documentclass{elsart3}

\def\l{\lambda}	         \def\d{\delta}					
\def\D{\Delta}

\def\us{\uparrow}         \def\ds{\downarrow}	
\def\dmu{\delta\mu}

\usepackage{epsfig}

\begin{document}

\begin{frontmatter}



\title{Spin Dependent Transport in Magnetic Nanostructures}

\author{S. Maekawa}

\address{Institute for Materials Research, Tohoku University, Sendai 980-8577, Japan }

\begin{abstract}
Recent progress in physics on spin dependent transport in magnetic
nanostructures is reviewed.  Special attention is paid on the spin
accumulation and spin current caused by spin injection into
non-magnetic metals and semiconductors and superconductors.  
A variety of phenomena induced in nano-superconductor/ferromagnet
devices are proposed, examining the spin-charge separation in superconductors.
\end{abstract}



\end{frontmatter}


\section{Introduction}

Discovery in 1988 of the so-called giant magnetoresistance (GMR) effect
in magnetic multilayers \cite{baibich,binash}
raised the curtain for an intensive new research
effort on magnetic materials and magnetic thin films which profited
enormously from the application of microfabrication techniques \cite{book}.  
Semiconductor devices have been the main actors on the electronics
stage during the latter half of twentieth century.  
The size of such devices has been getting steadily smaller following
the impressive progress in microfabrication techniques.  
In addition, the physics of such small semiconductor devices, i.e., 
mesoscopic physics, has emerged as a subject of study in its own right.  
It is also relevant to remember the considerable quantity of new physics 
which has emerged since the discovery of high temperature superconductors
in 1986.  In this latter field, one of the main concerns is the behavior
of spin and charge and their interactions.  On the other hand, to date
in semiconductor devices only the movement of charge has found application
with spin being usually considered as an irrelevant degree of freedom.  
The discovery of GMR signaled a starting point for the physics and 
application of the interaction between spin and charge and shed light 
to the spin polarized tunneling \cite{julliere,maekawa}
which was studied in advance of GMR.  
Here, I would like to review the new physics of spin dependent transport
in magnetic nanostructures.  Special attention is paid on the spin
accumulation and spin current caused by spin injection into non-magnetic
metals and semiconductors and superconductors.  Examining the spin-charge
separation in superconductors
\cite{schrieffer}, a variety of phenomena induced
in nano-superconductor/ferromanget devices are proposed.

\section{Spin Accumulation}

\begin{figure}
\epsfxsize=0.96\columnwidth			
\centerline{\hbox{\epsffile{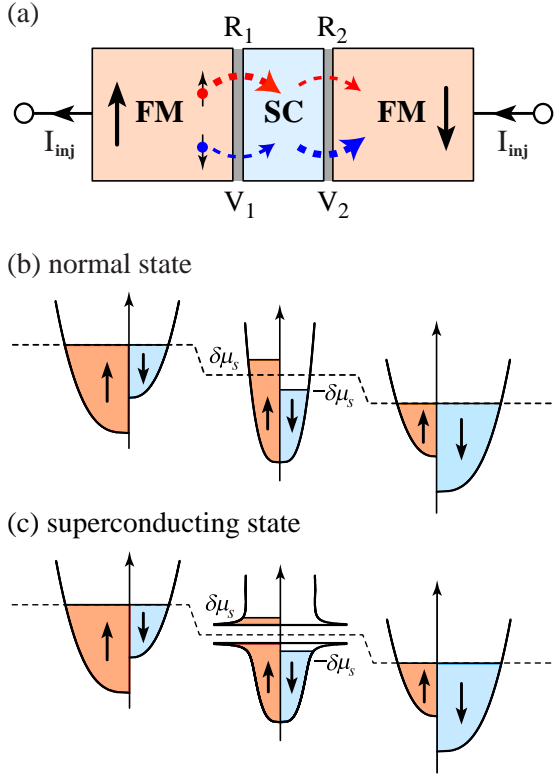}}}
\vskip 0.2cm
\caption{ (a) Double tunnel junction comprising two ferromagnets
(FM) and a superconductor (SC) separated by insulating barriers.
$R_1$ and $R_2$ are the tunnel resistances of the left and right
junctions with the voltage drops, $V_1$ and $V_2$, across the
barriers ($V=V_1+V_2$ is the total
voltage drop across the entire double tunnel junction).
Schematic diagram of the densities of states of FMs (left and light)
and SC (middle) in the antiparallel alignment of magnetizations of
FMs are shown when SC is in the normal state (b) and in the
superconducting state (c).
$\dmu_s$ denotes the shift in the chemical potentials of the spin subbands.
    }
\label{fig:fig1}                                       
\end{figure}

Let us consider a double tunnel junction device shown in Fig.~1 (a), 
where two ferromagnetic electrodes and a non-magnetic island are 
separated by two tunnel barriers.  In the device, the magnetic moments 
of two ferromagnetic electrodes are taken to be antiparallel and 
the electric current is introduced from the left electrode.  
In Fig. 1(b), the schematic electronic structure is presented.  
Here, electrons with up spin mainly come into the central island from 
the left electrode, whereas electrons with down spin mainly go out 
from the island to the right.  As a result, the spin imbalance occurs 
in the island with size smaller than the spin diffusion length 
($\l_s$), when electrons pass through it within the time less than
the spin relaxation 
time ($\tau_{s}$).  Since the magnetization is given by the difference
between the numbers of electrons with up and down spins times the 
Bohr magneton ($\mu_B$), the island is magnetized by the electric current.  
This is called spin accumulation.  Spin accumulation causes the tunnel
magnetoresistance (TMR) 
   \cite{barnas,brataas,imamura,takahashiPRL82,martinek,mattana}.  
In a double tunnel junction device with two
equivalent ferromagnetic electrodes, the spin polarization being $P$,
and two equal tunnel barriers, the TMR is given by 
  \begin{eqnarray}					
         TMR = P^2/(1 - P^2),            
     \label{eq:TMR}			
  \end{eqnarray}					
which is a half of that in the usual ferromagnetic tunnel junctions, i.e., 
$TMR = {2P^2/(1 - P^2)}$ 
  \cite{julliere,maekawa,moodera,miyazaki,slonczewski,parkin}.  
Therefore, the spin accumulation may be examined
by using TMR.  Recently, the TMR due to the spin accumulation has been studied
in GaMnAs/GaAs/GaMnAs double tunnel junction devices
  \cite{mattana}.  
\begin{figure}   					
\epsfxsize=0.94\columnwidth				
\centerline{\hbox{\epsffile{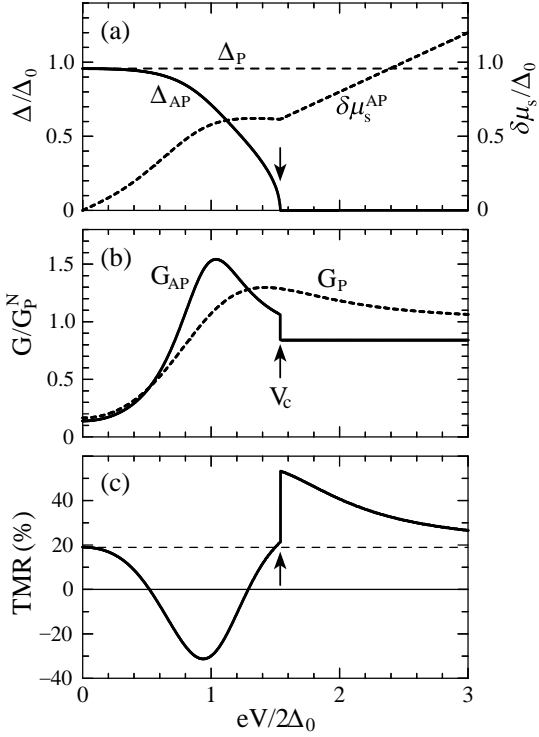}}}	
\vskip 0.2cm
\caption{                                          	
(a) Superconducting gaps, $\D_{P}$ and $\D_{AP}$,	
for the parallel (P) and antiparallel (AP) alignments
of magnetizations, respectively, and the half of the spin
splitting between the spin-up and spin-down bands
in the AP alignment, $\dmu_s^{AP}$ .	
(b) Tunnel conductances, $G_P$ and $G_{AP}$, for the	
P and AP alignments, respectively.			
(c) Tunnel magnetoresistance (TMR).			
Thin dashed line indicates TMR $= P^2/(1-P^2)$	
in the normal state.  The values $P=0.4$ and		
$T/T_c=0.5$ are taken for all curves.				
} 							
\label{fig:fig2}                                     
\end{figure}                                            
When the central island in Fig.~1 (a) becomes superconducting, the competition
between spin accumulation and superconductivity occurs as depicted in Fig.~1 (c).
In the superconducting state, there appears the superconducting gap ($\Delta$),
which competes with the spin accumulation.  In Fig.~2 (a), the theoretical
results of the shift of the Fermi level ($\d\mu_s$) due to spin accumulation
and $\Delta$ are presented as functions of bias voltage ($V$) in the double
tunnel junction device where the magnetic moments in the electrodes are
antiparallel.  
When the bias voltage becomes of the order of the gap, $eV/2 \sim \Delta/P$, 
the superconductivity is suppressed
     \cite{takahashiPRL82}. 
In Fig.~2 (b), the tunnel conductances, 
$G_{AP}$ and $G_P$, are shown in the device with antiparallel and parallel
magnetic moments, respectively.  For the antiparallel magnetic moments, 
the superconductivity is suppressed at $V = V_c$ and the conductance jumps.
The TMR in the device is given in Fig.~2 (c), where the dashed line shows
the value given by Eq.~(1).  We note that the TMR oscillates as a function
of $V$ since $\Delta$ depends on spin accumulation \cite{daibou,johansson,ono}.
The suppression of the superconducting gap by spin accumulation may be
observed by the superconducting critical current, $J_c$, following the relation, 
$J_c \propto \D^3$ 
   \cite{takahashiPRL82,vasko,dong,yeh99,liuPC01,chenPRL88}.  
\begin{figure}						
  \epsfxsize=0.84\columnwidth				
  \centerline{\hbox{\epsffile{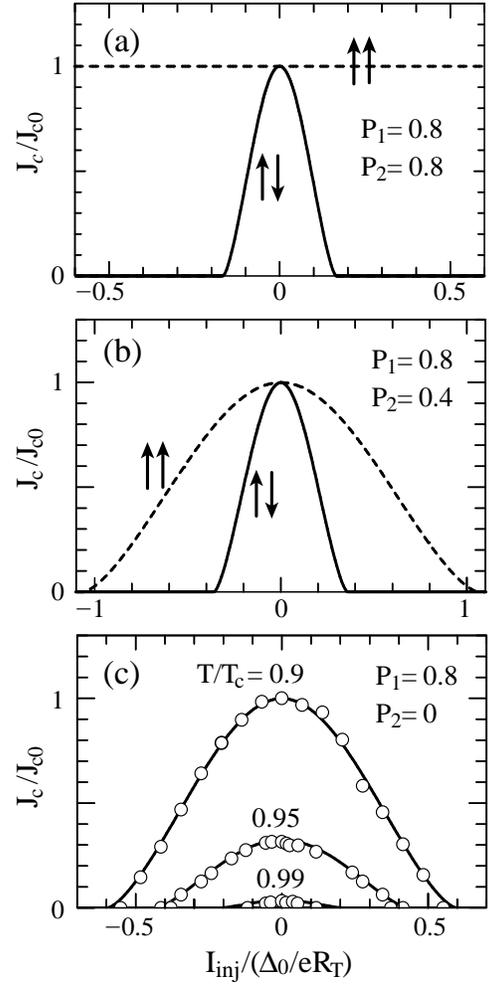}}}		
  \caption{						
Critical current ${J_c}$ as a function of injection	
current $I_{inj}$ for the spin polarizations $P_1$ of FM1 and	
different values of $P_2$ of FM2 at $T/T_c=0.9$:  				
(a) $P_1=0.8$ and $P_2=0.8$,  (b) $P_1=0.8$ and $P_2=0.4$,	
and (c) $P_1=0.8$ and $P_2=0$. 					
Open circles indicate the critical current at		
$T=80${\,}K, $84${\,}K and $87${\,}K ($T_c\sim 89${\,}K)	
in Au/YBa$_2$Cu$_3$O$_7$/LaAlO$_3$/Nd$_{2/3}$Sr$_{1/3}$MnO$_3$	
junctions \protect\cite{dong}.   				
   }							
  \label{fig3}						
\end{figure}						
In Fig.~3, $J_c$ normalized by that at zero temperature, $J_{c0}$,
is plotted as a function of the injected current, $I_{inj}$, at $T/T_c=0.9$,
the temperature ($T$) being normalized by the critical temperature ($T_c$).  
Here, the spin polarization of one of the ferromagnetic electrodes is taken
to be $P_1=0.8$ and that of the other one is chosen to be the values,
$P_2 = 0$, 0.4 and 0.8.
The tunnel resistance of the both tunnel barriers is equal and is
taken to be $R_{\rm T}$.  For the tunnel device with the same ferromagnetic
electrodes ($P_1 = P_2$),  the superconductivity is strongly suppressed
for the antiparallel magnetic moments, whereas it is not for the parallel
moments as seen in Fig.~3 (a).  
On the other hand, for the device with different ferromagnets ($P_1 \ne P_2$), 
the superconductivity is suppressed even for the parallel moments since spins
coming into and going out from the island is not equal (Fig.~3 (b)).  
When one of the electrodes is non-magnetic, $P_2$ is equal to zero (Fig.~3 (c)).  
The experimental result obtained in Nd$_{1-x}$Sr$_{x}$MnO$_3$/YBa$_2$Cu$_3$O$_7$/Au
devices 
  \cite{dong}
and theoretical ones are shown by circles and solid lines, respectively,
in Fig.~3 (c), where Nd$_{1-x}$Sr$_{x}$MnO$_3$ is a ferromagnetic
electrode and YBa$_2$Cu$_3$O$_7$ is a high temperature superconductor with
$T_c = 89$ K.  As noted above, the shift of the Fermi level ($\d\mu_s$)
due to spin accumulation
competes with the superconductivity, which disappears at 
$V_c \sim 2\D/eP$.  It is also known that the superconductivity is
suppressed by an applied magnetic field (Pauli paramagnetic effect).
In this case, the critical field is given by $H_c \sim 2\Delta/\mu_B$.  
By comparing these two effects, we find that the bias voltage of 0.1V 
corresponds to the magnetic fields of the order of $10^2$ T for 
superconductivity.  It is very hard to suppress the high temperature 
superconductivity by an applied magnetic field, since it is too large 
to access it in the usual experimental conditions.  On the other hand, 
the bias voltage of the order of 0.1V may be obtained easily.  
This fact suggests potentials for application of spin accumulation
in various fields.

\begin{figure}						
  \epsfxsize=0.98\columnwidth				
  \centerline{\hbox{\epsffile{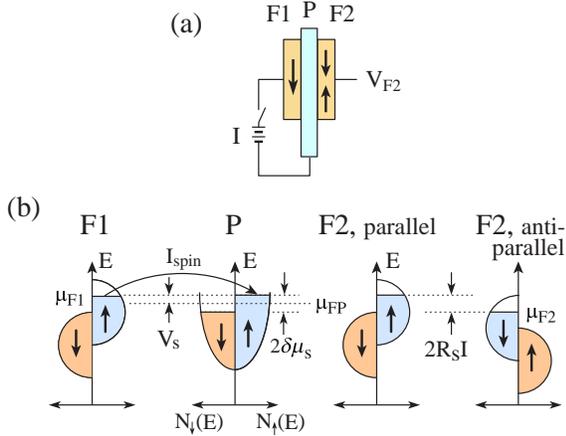}}}		
  \caption{  
(a) Three terminal device introduced by Johnson \cite{johnsonPRL70}.
Arrows in F1 and F2 refer to the magnetization orientation.
(b) Diagram of the densities of states of the 
ferromagnet(F1)/paramagnet(P)/ferromagnet(F2) system
in (a).					
   }  
\label{fig4}							
\end{figure}						

\section{Spin Injection Devices}

Johnson and Silsbee in 1985 \cite{johnson} and Johnson in 1993
  \cite{johnsonPRL70}
have proposed a three-terminal device shown in Fig.~4 (a), which consists
of a non-magnetic metal (P) and two ferromagnets (F1 and F2).
Let us consider that both F1 and F2 are half-metallic, for simplicity.  
Then, electrons with up spin are injected from F1 into 
P as seen in Fig.~4 (b).  As a result, the Fermi level with 
up spin in P is different from that with down spin.  
Therefore, the voltage ($V_2$) depends on whether the magnetic 
moment of F2 is paralle or antiparallel to that of F1
   \cite{valet,fert-lee,hershfield,johnson-bayers}. 
Recently, well-defined three-terminal devices were prepared 
by using the microfabrication technique \cite{jedema,george}.  
Here, it is important 
that the size of the devices is smaller than the spin diffusion length
($\l_{s}$), 
which is less than 1 $\mu$m in the usual non-magnetic metals.  
Since the signal in such spin injection devices is obtained as 
the voltage induced by the electric current, a low-carrier material 
in the nonmagnetic part will provide an opportunity for obtaining 
the larger signal.  Therefore, semiconductors such as GaAs with 
long spin diffusion length may be good candidates in the devices
   \cite{takahashiPRB67,fert-jaffres,rashba,schmidt}.  
We note that superconductors are low-carrier systems for spin, 
since spin is carried by quasi-particles whereas charge is by Cooper pairs.  
Thus, the large spin injection signal is also expected in the devices 
with superconductors \cite{takahashiPRB67}.

\section{Anomalous Hall Effect}

\begin{figure}						
  \epsfxsize=0.84\columnwidth				
  \centerline{\hbox{\epsffile{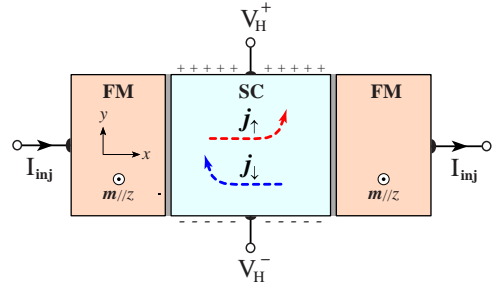}}}		
  \caption{  
Spin injection device of a double tunnel junction	
FM/SC/FM in the parallel magnetic moments of FMs,
in which a nonequilibrium Hall voltage $V_H$ is induced
in the transverse direction by injection of	
spin-polarized current.					
   }  
\label{fig5}							
\end{figure}						

Anomalous Hall effect in a ferromagnet is caused by spin polarized
 current scattered by the spin-orbit interaction at impurities
  \cite{hirschPRL99,zhang,otani,takahashiPRL88}.  
In other words, spin current shifts to charge current perpendicular 
to the applied bias voltage by the scattering \cite{hirschPRL99}.
Let us consider a double tunnel junction device in Fig.~5, where two 
electrodes are made of the same ferromagnet with magnetic moments 
perpendicular to the film plane ($z$-direction).  The current ($I_{inj}$) 
is injected from the left electrode to the non-magnetic central 
one ($x$-direction).  In this case, the charge and spin currents, 
respectively, are expressed as
  \begin{equation}					
I_{charge} = I_\us + I_\ds,                \\
  \end{equation}					
  \begin{equation}					
I_{spin}   = I_\us - I_\ds = PI_{inj},           
     \label{eq:Is}			
  \end{equation}					
where $I_\us$ and $I_\ds$ are the current with spin up and down, respectively, 
and $P$ is the spin polarization in the ferromagnetic electrodes.  
The Hall current is induced perpendicular to the injected current 
($y$-direction) and the Hall
voltage is given by
  \begin{eqnarray}					
      V_H  \propto
      {\mathop{z}^\rightarrow} \times {\mathop{I}^\rightarrow}_{spin}
      = {\mathop{z}^\rightarrow} \times P {\mathop{I}^\rightarrow}_{inj}.
     \label{eq:VH}			
  \end{eqnarray}					
It is interesting to see the Hall voltage in the superconducting state.  
As mentioned in the previous section, the charge current is carried by 
Cooper pairs and is not affected by impurities.  Since impurity 
scattering occurs only for quasi-particles which carry spin current, 
the Hall effect in superconductors proves that the anomalous Hall 
effect is due to spin current but not charge
  \cite{takahashiPRL88}.

\section{Conclusion}

Recent progress in microfabrication techniques has brought spin-electronics
devices of the order of or smaller than the spin diffusion length ($\l_{s}$)
 \cite{yamashita}, 
which provide a variety of spin dependent phenomena due to spin accumulation
and spin current.  The phenomena also depend on the geometry of the devices.
Spin dependent transport in magnetic nanostructures is based on the
physics of the interaction between spin and charge.  In this sense, 
the spin-charge separation of electrons is a starting point for the 
physics of spin-electronics.

\bigskip

\noindent{\bf Achnowledgements:}
\medskip

I would like to thank S. Takahashi, H. Imamura, T. Yamashita and J. Martinek for continuous collaboration in this field.  This work has been supported by a Grant-in-Aid for Scientific Research from MEXT and CREST.


\end{document}